\documentclass[aps,twocolumn,showpacs]{revtex4}
\usepackage{epsfig}
\usepackage{amsmath}

\def\avg#1{\langle#1\rangle}

\def\tr{\mbox{tr}}
\def\be{\begin{equation}}       \def\ee{\end{equation}}
\def\bea{\begin{eqnarray}}      \def\eea{\end{eqnarray}}

\begin{document}
\title{Dynamic generation of spin orbit coupling}
\author{Congjun Wu and Shou-Cheng Zhang}
\affiliation{Department of Physics, McCullough Building, Stanford
University, Stanford CA 94305-4045}

\begin{abstract}
Spin-orbit coupling plays an important role in determining the
properties of solids, and is crucial for spintronics device
applications. Conventional spin-orbit coupling arises
microscopically from relativistic effects described by the Dirac
equation, and is described as a single particle band effect. In
this work, we propose a new mechanism in which spin-orbit coupling
can be generated dynamically in strongly correlated,
non-relativistic systems as the result of fermi surface
instabilities in higher angular momentum channels. 
Various spin-orbit couplings can emerge in these new phases, and their
magnitudes can be continuously tuned by temperature or other
quantum parameters.
\end{abstract}
\maketitle
 
Most microscopic interactions in condensed matter physics can be
accurately described by non-relativistic physics. 
However, spin-orbit (SP) coupling is a notable exception, which arises 
from the relativistic Dirac equation of the electrons \cite{kittel1987}. 
The emerging science of spintronics makes crucial use of the
S-P coupling to manipulate electron spins by purely electric means. 
The proposed Datta-Das device \cite{datta1990} modulates the current
flow through the spin procession caused by the SP coupling. 
More recently, Murakami, Nagaosa and Zhang\cite{murakami2003,murakami2003a}
proposed a method of generating the dissipationless spin current
by applying an electric field in the $p$-doped semiconductors.
This effect and the similar proposal for the $n$-doped
semiconductors\cite{sinova2004} both make crucial use of the
SP coupling. In contrast to the generation of the spin
current by coupling to the ferromagnetic moment, purely electric
manipulation has an intrinsic advantage. However, unlike the
ferromagnetic moment, which can be spontaneously generated through
the strong correlation of spins, the conventional wisdom states
that the SP coupling is a non-interacting one-body effect,
whose microscopic magnitude is fixed by the underlying
relativistic physics.

On the other hand, recent interests are revived on the Landau-Pomeranchuk
(L-P) \cite{pomeranchuk1959} fermi surface instabilities, largely in 
connection with electronic liquid crystal states with spontaneously 
broken rotational symmetry \cite{halboth2000,oganesyan2001,barci2003,
metzner2003,kee2003,kee2003a}, and in connections with hidden orders 
in heavy fermion systems\cite{varma2003,palstra1985,ramirez1992}. 
Varma's recent work showed that the L-P instability could lead to 
the opening of an anisotropic gap at the fermi surface\cite{varma2003}.
In this paper, we show that the SP coupling can be generated dynamically
in a non-relativistic system through strong correlation effects as the
L-P instability in the spin channel with higher orbital angular momentum.
It emerges collectively after a phase transition, which is continuously
tunable either by temperature or by a quantum parameter at zero temperature.
Unlike the ferromagnet, our ordered phase keeps time reversal symmetry.
Also in contrast to the L-P instabilities considered by the
majority of previous theories, most translationally invariant
liquid phases in our model do not break rotational symmetry, and
some of them preserve time reversal and parity symmetries as well.
Most correlated phases in condensed matter physics are
characterized by their broken symmetries \cite{anderson1984}.
Solids break translational symmetry, liquid crystals break
rotational symmetry, superfluids and superconductors break gauge
symmetry and ferromagnets break time reversal symmetry and
rotational symmetry. As far as we are aware, the new phase
reported in this work is the only one besides the fermi liquid
which does not break any of the above symmetries. It is
distinguished from the fermi liquid by only breaking the
``relative spin orbit symmetry", a concept first introduced in the
context of the $^3$He liquid \cite{vollhardt1990}.

We first discuss the dynamic generation of SP coupling from the
L-P instability  within the Landau-Fermi liquid theory triggered by 
the negative Landau parameter $F_1^a$, and then present its exact
definition.
This instability lies in particle-hole channel with total spin one 
and relative orbital angular momentum one. 
Operators in matrix forms are defined as 
$Q^{\mu a}(r)= \psi^\dagger_\alpha(r) \sigma^\mu_{\alpha\beta} (-i
\hat\nabla^a ) \psi_\beta(r)$, where Greek indices denote the
direction in the spin space, Latin indices denote the direction in
the orbital space, and the operation of $\hat\nabla^a$ on the
plane wave is defined as $\hat\nabla^a  e^{i\vec{k}\cdot \vec{r}}=
(\nabla^a/|\nabla|) e^{i\vec{k}\cdot \vec{r}} =\hat k^a
e^{i\vec{k}\cdot \vec{r}}$. $Q^{\mu a}(r)$ is essentially the
spin-current operator up to a constant factor. We use a
Hamiltonian similar to that of Ref. \cite{oganesyan2001}, but in
the $F_1^a$ channel:
\bea  \label{HamL1S1}
H&=&\int d^3 {\vec r}
~\psi^\dagger_\alpha({\vec r}) (\epsilon(\vec{\nabla}) -\mu)
\psi_\alpha({\vec r})+ h_{\mu a} Q^{\mu a}(\vec r) \nonumber \\
&+& \frac{1}{2} \int d^3 {\vec r} d^3 {\vec r}^\prime ~f^a_1({\vec
r-\vec r^\prime)} Q^{\mu a}(\vec r) Q^{\mu a}(\vec r^\prime),
\eea 
where $\mu$ is the chemical potential and the small $h_{\mu a}$ is
dubbed as the ``spin-orbit field", which plays a role similar to
the external magnetic field. For later convenience
\cite{oganesyan2001}, we keep both the linear and the cubic terms
in the expansion of the single particle dispersion relation around
the fermi wavevector $k_f$, $\epsilon(\vec k)= v_f \Delta k [ 1+ b
(\Delta k/k_f)^2]$, with $\Delta k=k-k_f$. We assume that the
Fourier components of $f_1^a(\vec r)$ take the form $f_1^a(q)=\int
d \vec r ~ e^{i \vec q \vec r} f^a_1(r) = f^a_1/(1+\kappa |f^a_1|
q^2)$ and define the dimensionless Landau parameter $F^a_1=N_f
f^a_1$, where $N_f$ is the density of states at fermi energy. The
symmetry of the Hamiltonian (\ref{HamL1S1}) is a direct product
$SO(3)_L \otimes SO(3)_S$ in the orbital and spin channels.

We define the spin-orbit susceptibility  as $\chi_{\mu a,\nu
b}=\avg{Q_{\mu a}}/h_{\nu b}$ in the limit $h_{\nu b}\rightarrow
0$, which is diagonal, {\it i.e.} $\chi_{\mu a,\nu b}=\chi
\delta_{\mu a}\delta_{\nu b}$, in the normal fermi  liquid phase.
The fermi liquid correction to $\chi$ is given by \bea \chi_{FL}=
\chi_0 \frac{m^*}{m }\frac{1}{1+F^a_1/3}, \eea with  $m^*/m$ the
ratio between  the effective and  bare masses. The spin-orbital
susceptibility is enhanced for $F^a_1<0$  and divergent as the
critical point $F^a_1=-3$ is reached.

In the mean-field (MF) analysis, the p-h channel triplet order
parameter is defined as $n^{\mu a}(\vec r)=-\int d \vec r~
f^a_1(r-r^\prime) \avg{Q^{\mu a}(r^\prime)}$, and the external
spin-orbit field $h_{\mu a}$  is set to zero. Using the uniform
ansatz $n^{\mu a}(r)= n^{\mu a}$, Eq. \ref {HamL1S1} is decoupled into
$H_{MF}=\int d^3 {\vec r}
~\psi^\dagger({\vec r}) (\epsilon(\vec{\nabla}) -n^{\mu a}
\sigma^\mu (-i \hat\nabla^a) -\mu) \psi({\vec r}) 
+ Vn^{\mu a} n^{\mu a}/(2|f^a_1|),$
with  $V$  the space volume. 
The self-consistent equation for the order parameters  reads
\bea\label{selfconsistent}
n^{\mu a} &=&
|f^a_1| \int \frac{d^3 \vec k}{(2\pi)^3} \avg{ \psi^\dagger(k)
\sigma^\mu \hat k^a \psi(k)},
\eea
which is valid when the interaction range $r_0 = \sqrt{\kappa |f^a_1|}$
is much larger than the  distance between particles $1/k_f,$ {\it
i.e.}, the dimensionless parameter $\eta=k_f\sqrt{\kappa
|f^a_1|}\gg 1$.  

The phase structures can be determined from the Ginzburg-Landau (G-L) 
free energy, which is similar to the triplet pairing order parameter 
in the $^3$He system \cite{leggett1975,vollhardt1990}.
Under the independent
SO(3) rotations in the orbital and spin spaces $R_L$ and $R_S$,
$n^{\mu a}$ transforms as $n^{\mu a} \rightarrow R_{L,\mu\nu}
n^{\nu,b} R^{-1}_{S,ba}$. Furthermore, $n^{\mu a}$ is even under
the time-reversal but odd under the parity transformation. 
With these symmetry requirements, the G-L free
energy can be constructed up to the quartic order as
\bea\label{GL}
\hspace{-5mm}
F(n)&=& A~ \tr[n^T n] + B_1~(\tr [n^T n])^2+B_2~\tr[ (n^T n)^2]. \eea 
Compared with the complex order parameter in the superfluid $^3$He case, 
the reality of the $n_{\mu a}$  restricts the free energy 
to contain only two quartic terms. 
Explicitly, $\tr [n^T n]= n^{\mu a} n^{\mu a} =
\lambda_1^2 +\lambda_2^2+\lambda_3^2$, and $\tr [ (n^T n)^2]=
n^{\mu a} n^{\mu b}  n^{\nu a} n^{\nu b}=
\lambda_1^4+\lambda_2^4+\lambda_3^4$, where $\lambda_{1,2,3}^2$
are eigenvalues of $n^T n$. 
For $B_2<0$ or $B_2>0$, Eq. \ref{GL} favors the the structures of
$(\lambda_1^2,\lambda_2^2,\lambda_3^2)$ to be  proportional to
$(1,0,0)$ or $(1,1,1)$, respectively. We name them as $\alpha$ or
$\beta$-phases whose general order parameter matrix structures are
given by 
\bea\label{ABphase} 
n^{\mu a}&=& 
\left\{ \begin{array}{l}
\bar n ~\hat d_\mu \hat e_a  \mbox{~~~$\alpha$-phase, for $B_2<0$} \\
\bar n ~D_{\mu a} \mbox{~~~~$\beta$-phase, for $B_2>0$},
\end{array}
\right.
\eea
where $\hat d$ and $\hat e$ are two arbitrary unit
vectors in the spin and orbital space respectively, $D_{\mu a}$ is
any SO(3) rotation matrix, and $\bar n$ is a real number. In other
words, the correlation functions of operators $Q^{\mu a}$ acquire
a long range part in the ordered states 
\bea 
\avg{Q^{\mu a}(\vec r) Q^{\nu b}(\vec r^\prime)} \rightarrow
\delta_{\mu\nu}\delta_{ab} \frac{\bar n^2}{|f^a_1|^2}\times
\left\{ \begin{array}{c}
\hat d_\mu \hat e_a  \mbox{~$\alpha$-phase}   \\
D_{\mu a} \mbox{~~$\beta$-phase}
\end{array}
\right. 
\label{LRO}
\eea as $|\vec r-\vec r^\prime|\rightarrow
\infty$. 
This correlation function gives the rigorous definition for the new phases, 
independent of the approximate Fermi liquid theory used here.

\begin{figure}
\centering\epsfig{file=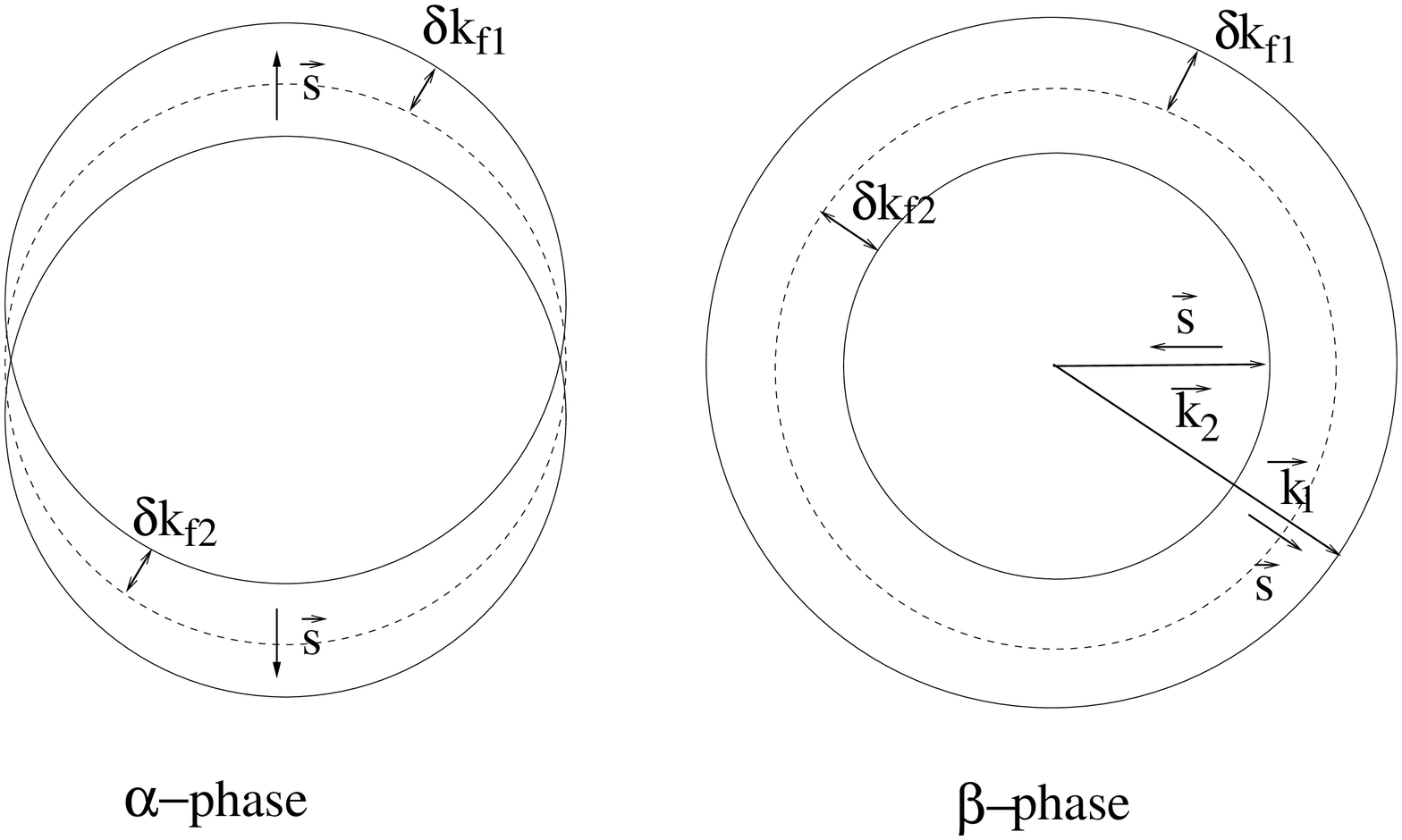,clip=1,width=\linewidth,angle=0}
\caption{The L-P instability in the $F^a_1$ channel,
with dashed lines marking the fermi surface before symmetry breaking.
The fermi surface distortion is anisotropic in the $\alpha$-phase,
while it is isotropic in the $\beta$-phase with dynamic generation
of spin-orbit coupling.
}\label{phasediagram}
\end{figure}

The $\alpha$-phase is a straightforward generalization of the nematic
fermi liquid \cite{oganesyan2001} to the triplet channel as shown
in Fig. \ref{phasediagram}, where the spin and orbital degrees of
freedom  remain decoupled, and the rotational symmetry is broken.
Taking a special case $n^{\mu a}=\bar n \delta_{\mu z} \delta_{a z}$, the
dispersion relations for  spin up and down branches are
$\xi^A(k)_{1,2}=\epsilon(k)-\mu\mp \bar n \cos\theta$
respectively, where $\theta$ is the angle between $k$ and $z$-axis.
The fermi surfaces for the two spin components are distorted in an
opposite way as $\Delta k_{f1,2}(\theta)/k_f=\pm x \cos\theta~ (1-
b x^2 \cos^2\theta)-1/3~ x^2$, with the dimensionless parameter
$x=\bar n/(v_f k_f)$. The chemical potential $\mu$  is shifted to
ensure the particle number conservation as  $\delta \mu /(v_f
k_f)=-x^2/3$. The remaining symmetry is $SO(2)_L\otimes SO(2)_S$
with the Goldstone manifold $S^2_L\otimes S^2_S$. Two Goldstone
modes are the oscillations of the distorted fermi  surfaces, and
the other two are the oscillations of the spin directions.

In the $\beta$-phase,  the rotational symmetry is preserved with the
dynamic generation of spin-orbit coupling as shown in Fig. \ref{phasediagram}.
For example, with the ansatz $n^{\mu a}=\bar n \delta_{\mu a}$,
the MF Hamiltonian is reduced to
\bea H_{MF}=
\sum_k \psi^\dagger(k) (\epsilon(k)-\mu -
\bar n \vec \sigma \cdot \hat k
) \psi(k). \label{BMF}
\eea
The single particle states can be classified according to the
eigenvalues $\pm 1$ of the helicity
operator $\vec \sigma\cdot \hat k$, with dispersion relations
$\xi^B(k)_{1,2}=\epsilon(k)-\mu\pm \bar n$. The fermi surface
distortions of two helicity bands are $ \Delta k_{f1,2}/k_f= \pm x
(1-b x^2)-x^2 $ and the chemical potential shift $\delta \mu/(v_f
k_f)=-x^2$. Similar to the superfluid $^3$He-B phase, the
$\beta$-phase is essentially isotropic.
The orbital $\vec  L$ and the spin $\vec S$
angular momenta are no longer separately conserved, but the total
angular momentum $\vec J= \vec L+\vec S=0$ remains conserved. The
Goldstone manifold is $[SO(3)_L\otimes SO(3)_S]/
SO(3)_{L+S}=SO(3)$ with three branches of Goldstone modes. For the
general case of $n_{\mu a}=\bar n D_{\mu a}$, it is equivalent to
a redefinition of spin operators as $S^\prime_\mu=S_\nu D_{\nu a}
\delta_{a\mu}$, thus fermi surface distortions remain isotropic
and $\vec J^\prime=\vec L +\vec S^\prime$ is conserved.
A similar symmetry breaking pattern also appears in the quantum  chromodynamics
where the two-flavor chiral symmetry $SU(2)_L \times SU(2)_R$
is broken into the diagonal $SU(2)_{L+R}$ \cite{wilczek1998}.
In that case, both $SU(2)_{L,R}$ are internal symmetries,
and thus there is no flavor-orbit coupling.

The coefficients of the G-L free energy Eq. \ref{GL} can be
microscopically derived from the MF theory as
\bea
A&=&\frac{1}{2}(\frac{1}{|f^a_1|}-\frac{ N_f}{3}),
~~~ B_1= \frac{N_f}{20 v_f^2 k_f^2}(1 +\frac{b}{3}),\nonumber \\
B_2&=& \frac{N_f}{30 v_f^2 k_f^2}
(\frac{-1}{3}+b),
\eea
where $b$ describes the cubic part of the dispersion $\epsilon(k)$, as
explained earlier.
With the definition of  $\delta=1/|F^a_1|-1/3$, the L-P instability
takes place at $\delta<0$ {\it i.e.}, $F^a_1<-3$.
For $b<1/3$, {\it i.e.} $B_2<0$, the $\alpha$-phase appears
with $|\bar n|^2= \frac {|A|}{2(B_1+B_2)} $.
For $b>1/3$, {\it i.e.} $B_2>0$, the $\beta$-phase is stabilized at with
$|\bar n|^2=\frac{|A|}{2(3 B_1+B_2)}$.
The largest fermi surface distortion $\Delta
k_{f,max}/k_f$ in the $\alpha$-phase is larger than the uniform one $\Delta
k_f/k_f$ in the $\beta$-phase, thus a large positive $b$ is helpful to the
$\beta$-phase.  However, we emphasize that this is only one of the options
to change the sign of $B_2$.

To apply the $\alpha$ and $\beta$ phases in the lattice system, we only need
replace the $SO(3)_L$ symmetry with the lattice  point group.
For example, for the simple cubic lattice,
we define $Q^{\mu a}= i\{ c^\dagger (\vec x) \sigma^\mu c(\vec x+\hat e_a)
-h.c.\}$ with $\hat e_a$  the base vector in the $a$-direction.
The unbroken symmetry is $O_L\otimes SO(3)_S$ where $O_L$ is the orbit lattice
octahedral group.
The mean field Hamiltonian for the $\beta$ phase reads
\bea
H_{MF}=\sum_{k} \psi^\dagger_k( \epsilon_{k}-\mu-
\bar n \sigma_\mu \sin k_a \delta_{\mu a}) \psi_k,
\eea
with lattice momentum $\vec k$ restricted in the first Brillouin zone.
The helicity structure for each $\vec k$ is aligned
along the direction of  $(\sin k_x, \sin k_y, \sin k_z)$,
which breaks the symmetry down to the combined octahedral rotation
in the orbit and spin space $O_{L+S}$.
As a real space analogy, the hexagonal non-colinear anti-ferromagnet
YMnO$_3$ \cite{andreev1980,munoz2000} has the spin order pattern
inside the unit cell which is also invariant under the combined
spin-orbit point group rotations.
The difference is that the spin order in the $\beta$ phase
lies in the momentum space and no spatial spin order exists each lattice site.
The lattice $\alpha$-phase was originally studied under
the name of ``spin-split state' by Hirsch \cite{hirsch1990}
to explain the  phase transition at  $T_N=311K$ in the Chromium system.

By reducing the space dimension to 2, the mean field Hamiltonian
for the $\beta$-phase reduce to the familiar
Rashba\cite{rashba1960} and Dressselhaus Hamiltonians
\cite{dresselhaus1955} in the 2D semiconductor  heterostructures. The
order parameter $n_{\mu a}$ is a $3\times 2$ matrix with
$\mu=x,y,z$ and $a=x,y$. Its third row of $\mu=z$ can be
transformed to zero by performing suitable SO(3) rotation on the
index $\mu$, thus we take $n_{\mu a}$ as a $2\times 2$ matrix. The
G-L free energy is also the same as in Eq. \ref{GL}, but with the
new coefficients \bea A&=&\frac{1}{2}( \frac{1}{|f^a_1|}-\frac{
N_f}{2}),~ B_1= \frac{N_f}{32 v_f^2 k_f^2 },~ B_2=\frac{b N_f}{8
v_f^2 k_f^2},~~ \eea and the L-P instability occurs at $F^a_1<-2$.
The $\alpha$ and $\beta$-phase structures are similar as before in
Eq. \ref{ABphase}. However,  there are two options in the
$\beta$-phase with $n^{\mu a}=\bar n D_{\mu a}$, where $D_{\mu a}$
is a $O(2)$ matrix. If $\det D=1$, then $J_z=L_z+S_z$ is
conserved. With the MF ansatz $n_{\mu a}=\bar n \epsilon_{\mu a}$,
we arrive at the Rashba-like  Hamiltonian \bea H_{R}=\int d^2 \vec
r ~ \psi^\dagger \Big \{ \epsilon(\vec\nabla)-\bar n
\epsilon_{z\mu a} \sigma^\mu (-i \hat\nabla^a) \Big \} \psi. \eea
If $\det D=-1$, $J_z$ is not conserved while the energy spectrum
and fermi surface distortions are still the same as the case of
$\det D=1$. With the MF ansatz $n_{\mu a}=\bar n~
\mbox{diag}\{1,-1\}$, we arrive at the 2D Dresselhaus-like
Hamiltonian as \bea H_{D}=\int d^2 \vec r~ \psi^\dagger  \Big \{
\epsilon(\vec\nabla)-\bar n [\sigma_x (-i \hat \nabla^x)- \sigma_y
(-i \hat \nabla^y)] \Big \}\psi. ~~~ \eea

If we generalize the mechanism of dynamical generation of the
spin-orbit coupling to the spin 3/2 fermionic system, an interesting
phase can be obtained which preserves all familiar symmetries
including the parity symmetry, breaking only the relative spin
orbit symmetry.
It has been recently shown that any generic model of spin 3/2 with local
interactions has an exact $SO(5)$ symmetry in the spin space\cite{wu2003}.
The four spin components form the spinor representation of the $SO(5)$ group.
Using the Dirac $\Gamma$ matrix defined there, the spin 3/2 Landau
interaction functions are classified into  the SO(5)'s scalar, vector
and tensor channels \cite{wu2003}:
\bea
f_{\alpha\beta,\gamma\delta}(\vec p,\vec p^\prime)&=& f^s(\vec
p,\vec p^\prime)+f^v(\vec p, \vec p^\prime)
(\Gamma^a/2)_{\alpha\beta}
(\Gamma^a/2)_{\gamma\delta} \nonumber \\
&+&f^t(\vec p,\vec p^\prime) (\Gamma^{ab}/2)_{\alpha\beta}
(\Gamma^{ab}/2)_{\gamma\delta}.
\eea
We further pick out its $L=2$ part of  the orbital angular momentum
in the spin 2 vector channel denoted as the $F^v_2$ channel.
We define operators
$Q^{\mu a}(\vec r)=\psi^\dagger_\alpha
(\vec r) \Gamma^\mu_{\alpha\beta} \hat d^a(\vec \nabla)\psi_\beta (\vec r)
~ (1\le \mu, a\le 5) $, where $\hat d^a(\vec \nabla)=
[{\sqrt 3} \hat \nabla_x  \hat \nabla_y,
{-\sqrt 3}\hat  \nabla_x  \hat \nabla_z,
{\sqrt 3} \hat \nabla_y   \hat \nabla_z,
-\frac{1}{2} (3 \hat \nabla_z^2- \hat \nabla^2),
\frac{\sqrt 3}{2} (\hat \nabla_x^2 - \hat \nabla_y^2)] $.
The model Hamiltonian is constructed as follows:
\bea\label{HamL2S2}
H&=&\int d^3 {\vec r} ~\psi^\dagger_\alpha({\vec r})
(\epsilon(\vec{\nabla})
-\mu) \psi_\alpha({\vec r}) \nonumber \\
&+& \frac{1}{2}
\int d^3 {\vec r} d^3 {\vec r}^\prime ~f^v_2({\vec r-\vec r^\prime)}
Q^{\mu a}(\vec r) Q^{\mu a}(\vec r^\prime),
\eea
with the symmetry of  $SO(3)_L\otimes SO(5)_S$.
The order parameter is defined as before
$n^{\mu a}(r)=-\int d \vec r^3 f^v_2(\vec r-\vec r^\prime)
\avg{Q^{\mu a}(\vec r^\prime)}$ and the L-P instability
occurs when $F^v_2= N_f f^v_2(q=0)<-5$.

The ordered phases after the L-P instability can also be
classified into two categories as before: the $\alpha$-phase
with anisotropic fermi surface distortions
and $\beta$-phase with spin-orbit
coupling. The detail phase structures are much more complicated
here. For example, the $\alpha$-phase has two non-equivalent
configurations because the $L=2$ channel fermi surface distortions
can be either uniaxial or biaxial. A comprehensive classification
of all the possible phases is quite involved and is deferred to a
future work. We focus here on the $\beta$-phase with the order
parameter structure $n^{\mu a}=\bar n \delta_{\mu a}$. In this
case, the MF Hamiltonian is reduced into $H_{MF}=\int d^3 \vec r
\psi^\dagger(\vec r) \Big \{ \epsilon(\nabla)-\mu-\bar n
\delta_{\mu a} \Gamma^\mu d^a(\hat \nabla)\Big\} \psi(\vec r)$.
From the relation between the $\Gamma^\mu$ matrices and the
quadratic form of spin 3/2 matrices ${\vec S}$
\cite{murakami2003a}, it can be easily recognized the
Luttinger-like Hamiltonian\cite{luttinger1956} \bea H_{L}=\int d^3
\vec r \psi^\dagger(\vec r) \Big\{\epsilon(\nabla) -\mu-\bar n (-i
\hat\nabla \cdot \vec S )^2  \Big\} \psi(\vec r), \eea which is
the standard model for the hole-doped III-V semiconductors. The
original symmetry $SO(3)_L\otimes SO(5)_S$ in Eq. \ref{HamL2S2} is
broken into $SO(3)_{L+S}$ with ten branches of Goldstone modes.
Different from the spin 1/2 case, the order parameter $n^{\mu a}$
here is even both under the time-reversal and parity
transformations.

It is remarkable that most standard spin-orbit systems,
including the familiar Rashba, Dresselhaus, and Luttinger-like
Hamiltonians can all be generated dynamically by our mechanism,
which can compliment the conventional mechanism based on
relativistic single particle physics. 
If a system has a small spin-orbit coupling to begin with,
it acts like a spin-orbit field $h_{\nu a}$ which rounds off
the phase transition discussed here.
One way to experimentally detect the dynamic generation of the
spin-orbit coupling is through the anomalous Hall effect (AHE),
which relies on the spin-orbit coupling. 
Therefore, detecting the AHE signal turning on at a phase transition 
directly demonstrates that the phase transition breaks the relative 
spin-orbit symmetry.
We propose to systematically search for these new phases in
$^3$He, ultra-cold atomic systems, semiconductors, heavy fermion
materials and ruthenates, both in experiments and in microscopic
numerical simulations. The new $\beta$ phases can be
experimentally detected through the standard predictions based on
the spin-orbit coupling, but would have the remarkable and
distinctive feature that these effects turn on and off near a
phase transition. We conjecture that besides the familiar
superfluid A and B-phases, $^3$He may contain the new phases
proposed in this work. The Landau parameter $F^a_1$ in $^3$He was
determined to be negative from various experiments
\cite{leggett1970,corruccini1971,osheroff1977,greywall1983} such
as the normal-state spin diffusion constant, spin-wave spectrum,
and the temperature dependence of the specific heat {\it etc}. It
varies from around $-0.5$ to $-1.2$ with increasing pressures to
the melting point, reasonably close to the instability point
$F^a_1=-3$. Even though we presented mean field descriptions of
the new phases with dynamically generated spin orbit coupling, the
existence of these phase can obviously be studied by exact
microscopic calculations of the correlation function Eq.
(\ref{LRO}) for realistic models.

We thank J. P. Hu and S. Kivelson for helpful discussions. CW
especially thanks E. Fradkin and V. Oganesyan for their education
on  nematic fermi liquids. This work is supported by the NSF under
grant numbers DMR-0342832, and the US Department of Energy, Office
of Basic Energy Sciences under contract DE-AC03-76SF00515. CW is
also supported by the Stanford Graduate Fellowship program.


\begin{thebibliography}{30}
\expandafter\ifx\csname natexlab\endcsname\relax\def\natexlab#1{#1}\fi
\expandafter\ifx\csname bibnamefont\endcsname\relax
  \def\bibnamefont#1{#1}\fi
\expandafter\ifx\csname bibfnamefont\endcsname\relax
  \def\bibfnamefont#1{#1}\fi
\expandafter\ifx\csname citenamefont\endcsname\relax
  \def\citenamefont#1{#1}\fi
\expandafter\ifx\csname url\endcsname\relax
  \def\url#1{\texttt{#1}}\fi
\expandafter\ifx\csname urlprefix\endcsname\relax\def\urlprefix{URL }\fi
\providecommand{\bibinfo}[2]{#2}
\providecommand{\eprint}[2][]{\url{#2}}

\bibitem[{\citenamefont{Kittel}(1987)}]{kittel1987}
\bibinfo{author}{\bibfnamefont{C.}~\bibnamefont{Kittel}},
  \emph{\bibinfo{title}{Quantum theory of solids}},
  \bibinfo{howpublished}{Generic} (\bibinfo{year}{1987}).

\bibitem[{\citenamefont{Datta and Das}(1990)}]{datta1990}
\bibinfo{author}{\bibfnamefont{S.}~\bibnamefont{Datta}} 
\bibnamefont{and }
  \bibinfo{author}{\bibfnamefont{B.}~\bibnamefont{Das}},
  \bibinfo{journal}{Appl. Phys.  Lett.} \textbf{\bibinfo{volume}{56}},
  \bibinfo{pages}{665} (\bibinfo{year}{1990}).

\bibitem[{\citenamefont{Murakami et~al.}(2003)\citenamefont{Murakami, Nagaosa,
  and Zhang}}]{murakami2003}
\bibinfo{author}{\bibfnamefont{S.}~\bibnamefont{Murakami}},
  \bibinfo{author}{\bibfnamefont{N.}~\bibnamefont{Nagaosa}}, \bibnamefont{and}
  \bibinfo{author}{\bibfnamefont{S.~C.} \bibnamefont{Zhang}},
  \bibinfo{journal}{Science} \textbf{\bibinfo{volume}{301}},
  \bibinfo{pages}{1348} (\bibinfo{year}{2003}).

\bibitem[{\citenamefont{Murakami et~al.}(unpublished)\citenamefont{Murakami,
  Nagaosa, and Zhang}}]{murakami2003a}
\bibinfo{author}{\bibfnamefont{S.}~\bibnamefont{Murakami, {\it et al} }},
\bibinfo{howpublished}{cond-mat/0310005} (\bibinfo{year}{unpublished}).

\bibitem[{\citenamefont{Sinova et~al.}(2004)\citenamefont{Sinova, Culcer, Niu,
  Sinitsyn, Jungwirth, and MacDonald}}]{sinova2004}
\bibinfo{author}{\bibfnamefont{J.}~\bibnamefont{Sinova, {\it et al}}},
\bibinfo{journal}{Phys. Rev. Lett.}
\textbf{\bibinfo{volume}{92}}, \bibinfo{pages}{126603}
(\bibinfo{year}{2004}).

\bibitem[{\citenamefont{Pomeranchuk}(1959)}]{pomeranchuk1959}
\bibinfo{author}{\bibfnamefont{I.~I.} \bibnamefont{Pomeranchuk}},
  \bibinfo{journal}{Sov. Phys. JETP.} \textbf{\bibinfo{volume}{8}},
  \bibinfo{pages}{361} (\bibinfo{year}{1959}).

\bibitem[{\citenamefont{Halboth and Metzner}(2000)}]{halboth2000}
\bibinfo{author}{\bibfnamefont{C.~J.} \bibnamefont{Halboth {\it et al} }}
  \bibinfo{journal}{Phys. Rev. Lett.} \textbf{\bibinfo{volume}{85}},
  \bibinfo{pages}{5162} (\bibinfo{year}{2000}).

\bibitem[{\citenamefont{Oganesyan et~al.}(2001)\citenamefont{Oganesyan,
  Kivelson, and Fradkin}}]{oganesyan2001}
\bibinfo{author}{\bibfnamefont{V.}~\bibnamefont{Oganesyan  {\it et al} }},
  \bibinfo{journal}{Phys. Rev. B} \textbf{\bibinfo{volume}{64}},
  \bibinfo{pages}{195109} (\bibinfo{year}{2001}).


\bibitem[{\citenamefont{Barci and Oxman}(2003)}]{barci2003}
\bibinfo{author}{\bibfnamefont{D.~G.} \bibnamefont{Barci {\it et al}}}
\bibinfo{journal}{Phys. Rev. B} \textbf{\bibinfo{volume}{67}},
\bibinfo{pages}{205108} (\bibinfo{year}{2003}).

\bibitem[{\citenamefont{Metzner et~al.}(2003)\citenamefont{Metzner, Rohe, and
  Andergassen}}]{metzner2003}
\bibinfo{author}{\bibfnamefont{W.}~\bibnamefont{Metzner {\it et al}}},
\bibinfo{journal}{Phys. Rev. Lett.} \textbf{\bibinfo{volume}{91}},
\bibinfo{pages}{66402} (\bibinfo{year}{2003}).

\bibitem[{\citenamefont{Kee}(2003)}]{kee2003}
\bibinfo{author}{\bibfnamefont{H.~Y.} \bibnamefont{Kee}},
  \bibinfo{journal}{Phys. Rev. B} \textbf{\bibinfo{volume}{67}},
  \bibinfo{pages}{73105} (\bibinfo{year}{2003}).

\bibitem[{\citenamefont{Kee et~al.}(2003)\citenamefont{Kee, Kim, and
  Chung}}]{kee2003a}
\bibinfo{author}{\bibfnamefont{H.~Y.} \bibnamefont{Kee {\it et al}}},
\bibinfo{journal}{Phys. Rev. B} \textbf{\bibinfo{volume}{68}},
\bibinfo{pages}{245109} (\bibinfo{year}{2003}).

\bibitem[{\citenamefont{Varma}(unpublised)}]{varma2003}
\bibinfo{author}{\bibfnamefont{C.~M.} \bibnamefont{Varma}},
  \bibinfo{howpublished}{cond-mat/0311145} (\bibinfo{year}{unpublised}).

\bibitem[{\citenamefont{Palstra et~al.}(1985)\citenamefont{Palstra, Menovsky,
  Vandenberg, Dirkmaat, Kes, Nieuwenhuys, and Mydosh}}]{palstra1985}
\bibinfo{author}{\bibfnamefont{T.~T.~M.} \bibnamefont{Palstra {\it et al}}},
\bibinfo{journal}{Phys. Rev. Lett.}
  \textbf{\bibinfo{volume}{55}}, \bibinfo{pages}{2727} (\bibinfo{year}{1985}).

\bibitem[{\citenamefont{Ramirez et~al.}(1992)\citenamefont{Ramirez, Coleman,
  Chandra, Bruck, Menovsky, Fisk, and Bucher}}]{ramirez1992}
\bibinfo{author}{\bibfnamefont{A.~P.} \bibnamefont{Ramirez {\it et al}}},
  \bibinfo{journal}{Phys. Rev. Lett.} \textbf{\bibinfo{volume}{68}},
  \bibinfo{pages}{2680} (\bibinfo{year}{1992}).

\bibitem[{\citenamefont{Anderson}(1984)}]{anderson1984}
\bibinfo{author}{\bibfnamefont{P.~W.} \bibnamefont{Anderson}},
  \emph{\bibinfo{title}{Basic notations of condensed matter physics}}
  (\bibinfo{publisher}{ Westview Press },
  \bibinfo{year}{1994}).

\bibitem[{\citenamefont{Vollhardt and Wolfle}(1990)}]{vollhardt1990}
\bibinfo{author}{\bibfnamefont{D.}~\bibnamefont{Vollhardt}} \bibnamefont{and}
  \bibinfo{author}{\bibfnamefont{P.}~\bibnamefont{Wolfle}},
  \emph{\bibinfo{title}{The superfluid phases of helium 3}}
  (\bibinfo{publisher}{Taylor \& Francis}, \bibinfo{year}{1990}).

\bibitem[{\citenamefont{Leggett}(1975)}]{leggett1975}
\bibinfo{author}{\bibfnamefont{A.~J.} \bibnamefont{Leggett}},
  \bibinfo{journal}{Rev.  Mod.  Phys.} \textbf{\bibinfo{volume}{47}},
  \bibinfo{pages}{331} (\bibinfo{year}{1975}).

\bibitem[{\citenamefont{Wilczek}(1998)}]{wilczek1998}
\bibinfo{author}{\bibfnamefont{F.}~\bibnamefont{Wilczek}},
  \bibinfo{journal}{Nuclear Physics A} \textbf{\bibinfo{volume}{642}},
  \bibinfo{pages}{1C} (\bibinfo{year}{1998}).

\bibitem[{\citenamefont{Andreev and Marchenko}(1980)}]{andreev1980}
\bibinfo{author}{\bibfnamefont{A.~F.} \bibnamefont{Andreev {\it et al}}},
\bibinfo{howpublished}{Sov. Phys. Usp. {\bf 23}} (\bibinfo{year}{1980}).

\bibitem[{\citenamefont{Munoz et~al.}(2000)\citenamefont{Munoz, Alonso,
  Martinez-Lope, Casais, Martinez, and Fernandez-Diaz}}]{munoz2000}
\bibinfo{author}{\bibfnamefont{A.}~\bibnamefont{Munoz {\it et al} }},
\bibinfo{journal}{Phys. Rev. B}
 \textbf{\bibinfo{volume}{62}}, \bibinfo{pages}{9498} (\bibinfo{year}{2000}).

\bibitem[{\citenamefont{Hirsch}(1990)}]{hirsch1990}
\bibinfo{author}{\bibfnamefont{J.~E.} \bibnamefont{Hirsch}},
  \bibinfo{journal}{Phys. Rev. B} \textbf{\bibinfo{volume}{41}},
  \bibinfo{pages}{6820} (\bibinfo{year}{1990}).

\bibitem[{\citenamefont{Rashba}(1960)}]{rashba1960}
\bibinfo{author}{\bibfnamefont{E.~I.} \bibnamefont{Rashba}},
  \bibinfo{journal}{Soviet Physics-Solid State} \textbf{\bibinfo{volume}{2}},
  \bibinfo{pages}{1109} (\bibinfo{year}{1960}).

\bibitem[{\citenamefont{Dresselhaus}(1955)}]{dresselhaus1955}
\bibinfo{author}{\bibfnamefont{G.}~\bibnamefont{Dresselhaus}},
  \bibinfo{journal}{Phys. Rev.} \textbf{\bibinfo{volume}{100}},
  \bibinfo{pages}{580} (\bibinfo{year}{1955}).

\bibitem[{\citenamefont{Wu et~al.}(2003)\citenamefont{Wu, Hu, and
  Zhang}}]{wu2003}
\bibinfo{author}{\bibfnamefont{C.~J.} \bibnamefont{Wu {\it et al}}},
   \bibinfo{journal}{Phys. Rev. Lett.} \textbf{\bibinfo{volume}{91}},
  \bibinfo{pages}{186402} (\bibinfo{year}{2003}).

\bibitem[{\citenamefont{Luttinger}(1956)}]{luttinger1956}
\bibinfo{author}{\bibfnamefont{J.~M.} \bibnamefont{Luttinger}},
  \bibinfo{journal}{Phys. Rev.} \textbf{\bibinfo{volume}{102}},
  \bibinfo{pages}{1030} (\bibinfo{year}{1956}).

\bibitem[{\citenamefont{Leggett}(1970)}]{leggett1970}
\bibinfo{author}{\bibfnamefont{A.~J.} \bibnamefont{Leggett}},
  \bibinfo{journal}{J. Phys. C}
  \textbf{\bibinfo{volume}{3}}, \bibinfo{pages}{448} (\bibinfo{year}{1970}).

\bibitem[{\citenamefont{Corruccini et~al.}(1971)\citenamefont{Corruccini,
  Osheroff, Lee, and Richards}}]{corruccini1971}
\bibinfo{author}{\bibfnamefont{L.~R.} \bibnamefont{Corruccini {\it et al}}},
  \bibinfo{journal}{Phys. Rev. Lett.} \textbf{\bibinfo{volume}{27}},
  \bibinfo{pages}{650} (\bibinfo{year}{1971}).

\bibitem[{\citenamefont{Osheroff}(1977)}]{osheroff1977}
\bibinfo{author}{\bibfnamefont{D.~D.} \bibnamefont{Osheroff }},
  \bibinfo{journal}{Physica B \& C} \textbf{\bibinfo{volume}{90}},
  \bibinfo{pages}{20} (\bibinfo{year}{1977}).

\bibitem[{\citenamefont{Greywall}(1983)}]{greywall1983}
\bibinfo{author}{\bibfnamefont{D.~S.} \bibnamefont{Greywall}},
  \bibinfo{journal}{Phys. Rev. B} \textbf{\bibinfo{volume}{27}},
  \bibinfo{pages}{2747} (\bibinfo{year}{1983}).

\end{thebibliography}

\end{document}